\documentclass[sigconf]{acmart}
\usepackage{multirow}

\AtBeginDocument{%
  \providecommand\BibTeX{{%
    \normalfont B\kern-0.5em{\scshape i\kern-0.25em b}\kern-0.8em\TeX}}}

\copyrightyear{2021}
\acmYear{2021}
\setcopyright{rightsretained}
\acmConference[CUI '21]{CUI 2021 - 3rd Conference on Conversational User Interfaces}{July 27--29, 2021}{Bilbao (online), AA, Spain}
\acmBooktitle{CUI 2021 - 3rd Conference on Conversational User Interfaces (CUI '21), July 27--29, 2021, Bilbao (online), AA, Spain}
\acmDOI{10.1145/3469595.3469618}
\acmISBN{978-1-4503-8998-3/21/07}

\begin{document}

\title{Eliciting Spoken Interruptions to Inform Proactive Speech Agent Design}

\author{Justin Edwards}
\affiliation{%
  \institution{University College Dublin}
  \city{Dublin}
  \country{Ireland}}
\orcid{0000-0003-1487-9207}
\email{justin.edwards@ucdconnect.ie}

\author{Christian P. Janssen}
\affiliation{%
  \institution{Utrecht University}
  \city{Utrecht}
  \country{The Netherlands}}
\orcid{0000-0002-9849-404X}
\email{C.P.Janssen@uu.nl}

\author{Sandy J. J. Gould}
\affiliation{%
  \institution{Cardiff University}
  \city{Cardiff}
  \country{UK}}
\orcid{0000-0003-0476-4270}  
\email{goulda5@cardiff.ac.uk}

\author{Benjamin R. Cowan}
\affiliation{%
  \institution{University College Dublin}
  \city{Dublin}
  \country{Ireland}}
\orcid{0000-0002-8595-8132}
\email{benjamin.cowan@ucd.ie}

\renewcommand{\shortauthors}{Edwards, et al.}

\begin{abstract}
 Current speech agent interactions are typically user-initiated, limiting the interactions they can deliver. Future functionality will require agents to be proactive, sometimes interrupting users. Little is known about how these spoken interruptions should be designed, especially in urgent interruption contexts. We look to inform design of proactive agent interruptions through investigating how people interrupt others engaged in complex tasks. We therefore developed a new technique to elicit human spoken interruptions of people engaged in other tasks. We found that people interrupted sooner when interruptions were urgent. Some participants used access rituals to forewarn interruptions, but most rarely used them. People balanced speed and accuracy in timing interruptions, often using cues from the task they interrupted. People also varied phrasing and delivery of interruptions to reflect urgency. We discuss how our findings can inform speech agent design and how our paradigm can help gain insight into human interruptions in new contexts. 
        
\end{abstract}



\maketitle

\section{Introduction}

Speech interaction with agents is now commonplace. Current speech interaction methods  are limited, mostly using  wake words such as “Hey Google” or “Alexa” to commence interaction.  This constrains the types of interactions these systems can deliver. Future functionality such as giving users notifications or even initiating collaboration on tasks will need agents to be more proactive, interrupting users who may be engaged in other tasks. Recent work has begun to explore within what context speech agents may be able to interrupt \cite{cha_hello_2019}, yet we currently do not know how these interruptions should be designed, especially in contexts where this information may be urgent or time sensitive. Similar to other speech technology work \cite{gilmartin_social_2017, landesberger_urgent_2020, edlund_pause_2009}, our study aims to gather insight from human-human interaction to inform speech technology design. Specifically we look to identify how to design proactive agent interruptions through through a mixed-methods analysis of how people interrupt others when they are busy conducting a complex task.  To do this we develop a new technique to elicit human spoken interruptions of people actively engaged in another task, and from this seek to investigate what verbal behaviors interrupters engage in to get the attention of people engaged in other tasks. Our work shows that the level of urgency significantly affects how long it takes for people to start interrupting, with people interrupting faster with an urgent request. Linguistically, we found no quantitative effect of urgency on the use of access rituals, yet some participants used these access rituals consistently to forewarn interruptions. Our qualitative findings also show that there were a wide variety of strategies used by participants to time their interruptions, balancing speed and accuracy, with many stating that they waited for points of perceived low load to engage users in conversation. People also mentioned that they varied their prosody or word choice to convey the urgency of  messages when interrupting.  

\section{Related Work}
\subsection{Interruptions and Multitasking}

Interruptions are a common topic of study in human-computer interaction (HCI). While interrupting a task risks distraction, they may also bring benefits to productivity or facilitate a response to emergent tasks \cite{janssen_integrating_2015}. Interruptions are frequently studied in the form of notifications, which trigger task switches \cite{iqbal_investigating_2005, mcfarlane_comparison_2002}, and as self-interruptions, in which task switching is self-triggered \cite{dabbish_why_2011}. Critical to the study of interruptions is the observation of task switching between a main task (termed the primary task) and an interrupting task (termed the secondary task), with multitasking and interruptions being understood as a singular phenomenon on a continuum of time between these task switches \cite{salvucci_threaded_2008}.

When paying attention to interruptions, people tend to consider the impact of engaging in a secondary task on their primary task, balancing speed in completing tasks with avoiding errors in the primary task (known as the speed-accuracy tradeoff) \cite{brumby_recovering_2013}.  Although such a trade-off is considered, research on interruptions during driving shows that people tend to prioritize speed as a default strategy, interrupting a primary task as quickly as possible \cite{horrey_driver-initiated_2009}, having to be told to emphasise accuracy before it is prioritized \cite{brumby_fast_2011}. People also tend to time interruptions based on the status of the primary task, focusing on natural breakpoints within the task they are conducting. Natural breakpoints are moments between the end of one subtask and the beginning of the next. These tend to mark low-cost moments of interruption, that people are naturally good at coordinating, especially for self-interruption \cite{borst_problem_2010, janssen_integrating_2015, bailey_understanding_2008}. Examples of natural breakpoints include finishing typing a sentence in an email or the moment after turning on the toaster in the task of making breakfast. These breakpoints, although useful for tasks that can be broken down into clear discrete units,  are difficult for people to identify when tasks are continuous (i.e. when tasks are not reducible into discrete units of ongoing activities that do not overlap (see \cite{kieras_modern_2000}). Complex continuous tasks that may not have clear natural breakpoints are difficult to model in terms of ideal interruption moments \cite{semmens_is_2019} making it difficult to design interruptions for these tasks.  In these cases, forewarned (i.e. interruptions that come after a warning message) or negotiated interruptions (i.e. interruptions that offer a person a choice to postpone interruption) can allow people to prepare or select the best moment to engage in a secondary task,  leading to better primary task performance. They also allow people to better prepare for interrupting tasks when engaged in a complex continuous task like playing a video game and monitoring handover requests in autonomous driving \cite{mcfarlane_comparison_2002, van_der_heiden_priming_2017}.

\subsection{Multitasking, Speech Interfaces, and Interaction Initiation}
Interacting with speech interfaces can be an effective way to accomplish other tasks while otherwise engaged in a primary task \cite{luger_like_2016}. Speech interfaces have been shown to effectively support the execution of complex tasks like preparing a presentation without dangerously interfering with driving \cite{martelaro_exploration_2019}.  However, speech-based multitasking is more suitable for particular primary tasks, such as those that do not also involve the production of language \cite{edwards_multitasking_2019}. Multitasking with speech interfaces while driving has been a particularly popular area of research, with a 2017 meta-review of 43 studies of voice-recognition systems in the car noting that these systems impose some penalty on driving performance, but less so than visual-manual interfaces \cite{simmons_meta-analysis_2017}.  

Work thus far has focused on user-initiated task switches, rather than systems with mixed initiative. Recent work has begun to explore the contexts in which more proactive interruption by speech interfaces may be possible \cite{cha_hello_2019}. The work found that, when in the home, opportune moments for interruption are governed by aspects such as user busyness, primary task difficulty, the extent to which the primary task is repetitive, as well as a  person’s social availability and mood \cite{cha_hello_2019}. Seminal work on mixed-initiative interactions has also outlined ways of initiating proactive interactions more generally, emphasising social norms and attributes from human-human interaction, such as appropriate levels of formality in address, should be considered in the design of proactive agents \cite{horvitz_principles_1999, horrey_driver-initiated_2009}. Currently though there is little understanding as to how these proactive interruptions should be designed as spoken interactions, in terms of content and delivery. 

\subsection{Access Rituals and Urgent Speech}
One promising avenue for the design of speech based proactive interruptions is through the use of access rituals. Access rituals are short verbal and nonverbal behaviors people engage in at the beginning of or the end of an interaction with another person, signalling a request for or a ceding of access to that person \cite{goffman_relations_1971}. In the context of beginning a conversation, like what occurs during a spoken interruption, people tend to use a  number of common access rituals to initiate interaction \cite{krivonos_initiating_1975}, including verbal behaviors such as verbal salutes (e.g. “hi”), use of names or nicknames, or apologizers (e.g. “sorry” or “excuse me”). Access rituals have thus far been studied only in  situations where conversing with a partner is the only task, with little being known about how people interrupt others engaged in another task for the purpose of a conversation. 

One characteristic that may play an important role when interrupting a person through speech is interruption urgency. Although not focused on interruptions, recent work on speech agents shows that users’ speech signal varies with the urgency of the message they need to convey to an agent. Urgent speech varies from normal speech when interacting with an agent, leading to changes in prosody (i.e. the way speech sounds, acoustically and subjectively), most notably an increase in pitch, speaking rate, and intensity \cite{landesberger_urgent_2020, landesberger_what_2020}. Urgent speech also tends to be distinctive in semantics (i.e. the meanings of words) when compared to non-urgent speech, with some words being perceived as more urgent than other words independent of how they are delivered prosodically \cite{hellier_perceived_2002}.  When manipulating urgency, these studies tend to use a  gamified approach whereby rewards are altered to make urgent trials more high-stakes \cite{landesberger_urgent_2020, landesberger_what_2020}. This approach has been shown to be effective,  with participants producing speech in urgent trials that differs significantly from their speech in non-urgent trials \cite{landesberger_urgent_2020}. Urgent notifications also lead people to be more open to being interrupted \cite{vastenburg_considerate_2008}. This suggests that urgency may be a potentially important variable in the design of spoken interruptions.

\subsection{The Current Study}
Currently little is known about how people use speech to interrupt those who are busy conducting another complex task. It is thus difficult for proactive speech agent designers to identify ways in which these agents can approach interrupting otherwise engaged users to commence collaboration or relay important information.  Combining knowledge of interruptions, access rituals, and urgent speech, this work uses a mixed-methods approach to  explore how people interrupt others in order to inform proactive and mixed initiative speech agent design.  We contribute to this aim by  1) proposing a paradigm for eliciting spoken interruptions and observing their temporal and linguistic characteristics, using the game of Tetris as well as 2) quantifying and identifying the nuance of strategies that people use to interrupt people actively engaged in another task to engage them in conversation, in both urgent and non-urgent conditions. We use an experimental design that observes spoken interruptions in which one person interrupts another person who is engaged in another task. We use videos and audio recordings of the human Tetris player to control for Tetris task performance and reactions to interruptions. By casting human participants in the role of an interrupter, we seek to better understand spoken interruptions through a mixed-methods study of  of both \textit{when} and \textit{in what way} people interrupt other people using speech, as to inform proactive agent design. Based on the work summarized above we hypothesize that urgency will have a statistically significant effect on the time it takes to initiate an interruption (interruption onset) (H1) and how long an interruption lasts (interruption duration) (H2). We also hypothesise that use of access rituals will statistically significantly vary dependent on the urgency of the interruption (H3).  Through our qualitative data, we also aim to more deeply explore the various approaches our participants used speech to interrupt people engaged in another task. 

\section{Experimental Method}
\subsection{Participants}
52 crowdworkers (26 women, 24 men, 2 preferred not to specify; M\textsubscript{age} = 29.4 years, SD = 7.9 years) were recruited from a crowdsourcing platform (Amazon Mechanical Turk). All participants were native or near-native English speakers. Participants were all familiar with the game Tetris, with most indicating that either they had played before, but do not play regularly (N = 44; 84.6\% of sample) or that they play regularly (N = 3, 5.7\% of sample) (5 point Likert scale; 1 = I am not at all familiar with Tetris;  5 = I regularly play Tetris). The study took approximately 20 minutes and participants were paid  \$10 Mechanical Turk credit for participating in the research. The study received ethical approval through the university’s ethics procedures for low risk projects (Ethics code: HS-E-20-161).

\subsection{Materials}
\subsubsection{Tetris Interruptions Paradigm Rationale}
In our experiment, we sought to explore how people interrupt a partner when they are executing a primary task that requires ongoing attention and cannot be arbitrarily suspended (continuous) and allows for a broad variety of responses rather than a single fixed response (complex) \cite{kieras_modern_2000, salvucci_multitasking_2005}. We therefore devised an experimental paradigm around Tetris as the primary task. We chose Tetris as a  primary task as it has been shown it to be a “manageabley complex” task \cite{lindstedt_distinguishing_2019}: a task that has a variety of features to which someone must adapt, and which has a variety of structures of events, lending itself to a different adaptation strategies for different players. The paradigm was designed to ensure that the interaction context could believably be conducted online. Participants were told that they would be interacting with a remote partner who would be playing Tetris online and that they would have to deliver spoken interruptions to this partner. Further details of the paradigm design are outlined below.

\subsubsection{Tetris Task}
The trials within the paradigm used recorded, rather than live, Tetris gameplay. This means that the materials can be standardized across all participants so as to control for potential variability between the stimuli (e.g. variability within Tetris players and Tetris game states). That said, in order to maintain engagement and to elicit interruptions reflective of how people interrupt other people, participants were told that the pre-recorded videos were a live feed of a person playing Tetris. Participants were told that they were matched with a person who was currently playing Tetris at the start of the experiment. The experiment involved 2 practice trials followed by 16 experimental trials. These trials were generated from 3 minute videos of actual Tetris gameplay conducted by the lead author. Each trial was chosen to ensure that the game state reflected one in which the Tetris player was not at risk of losing when the interruption occurred. Specifically: 1) a Tetris game piece started at the top of the game board; 2) there were at least two rows and no more than half of the rows of the board which already contained Tetris pieces and 3) the falling speed of the game piece was set to the game minimum of 1.25 rows per second. Each trial was presented as a video on a webpage. Videos included a Tetris board and a box in the upper right corner indicating the next piece. Videos were presented at an 800x800 resolution, in color, on a neutral background, and without sound.

\subsubsection{Interrupting Task}
Participants were tasked with completing a set of interrupting tasks, requesting information from the Tetris player, similar to other interruptions research \cite{kubose_effects_2006}. Once a trial had started, a message would appear on-screen instructing the participant that they needed to request a certain piece of information from their partner. Messages appeared in large black font in a single line on the screen directly below the Tetris video after a random delay between 5000 and 15000 milliseconds. In each trial, participants were told what information they needed to request from their partner. To encourage naturalistic generation of utterances, the messages instructing participants on what to ask their partner included only key words rather than full, grammatically complete questions. Specifically, these messages instructed participants to “in your own words, ask your partner:'' followed by keywords. This was to ensure that participants were not led to read aloud or directly use the question prompt when forming their interruption utterance. Building on methods from previous research \cite{wu_see_2020,wu_mental_2020} we use keyword prompts rather than verbatim written instructions. This was to ensure that participants had to plan and generate utterances rather than directly replicating the task prompt. The prompt was displayed during the trial to eliminate confounds of task retrieval from memory on interruption planning. 

{\setlength{\tabcolsep}{4em}
\begin{table*}
\centering
\caption{Table of interruption prompts.}
\label{tab:prompts}
\begin{tabular}{|c|c|} 
\hline
\multicolumn{2}{|c|}{\textbf{Interruption prompts: “In your own words, ask partner: \_\_\_\_”} } \\ 
\hline
which hand using & last movie watched \\ 
\hline
any pets & favorite ice cream flavour \\ 
\hline
weather & what breakfast this morning \\ 
\hline
bed time last night & been to paris \\ 
\hline
age & favorite fruit \\ 
\hline
last series watched & favorite color \\ 
\hline
any siblings & lucky number \\ 
\hline
what dinner last night & keyboard color \\
\hline
\end{tabular}
\end{table*}
}

Questions focused on requesting details about their partner (task prompts are included in Table \ref{tab:prompts}). These were used for two reasons. Firstly, participants would not know the answers to these questions and thus would not be tempted to answer on their partner’s behalf. Secondly, these questions would all be of similarly low difficulty for their partner to answer. This meant the responses could believably be generated after a uniformly short delay, enhancing the realism of the paradigm. It also reduces any variance in question asking that may result from participants’ beliefs about question difficulty. 

\subsubsection{Partner’s Rating of Performance}
To keep participants engaged with the task they would interrupt (the Tetris game), participants were told “After each round is finished, your partner will be asked to rate how well you did in terms of how disruptive your question was. Your partner will be asked how much they agree with the following two statements: ‘My partner's question came at a good moment.’ and ‘My partner's question did not distract me.’” Participants were told that these ratings determined a final score and that the participant with the highest total score at the end of the experiment would receive a bonus reward. 

\subsubsection{Simulation of Player Responses to Interrupting Task}
Pre-recorded responses were used to answer the questions posed by the participant. These responses were recorded by a male and female member of the research team who were native speakers of Hiberno-English. The gender of the Tetris player was randomly assigned and balanced across participant gender. Responses were scripted to ensure that they were identical in content and structure. To enhance believability, recordings were made on built-in laptop microphones so audio quality is clear without being unexpectedly high-fidelity.  

\subsection{Experiment Conditions}
\begin{figure*}[h]
  \centering
  \includegraphics[width=0.5\linewidth]{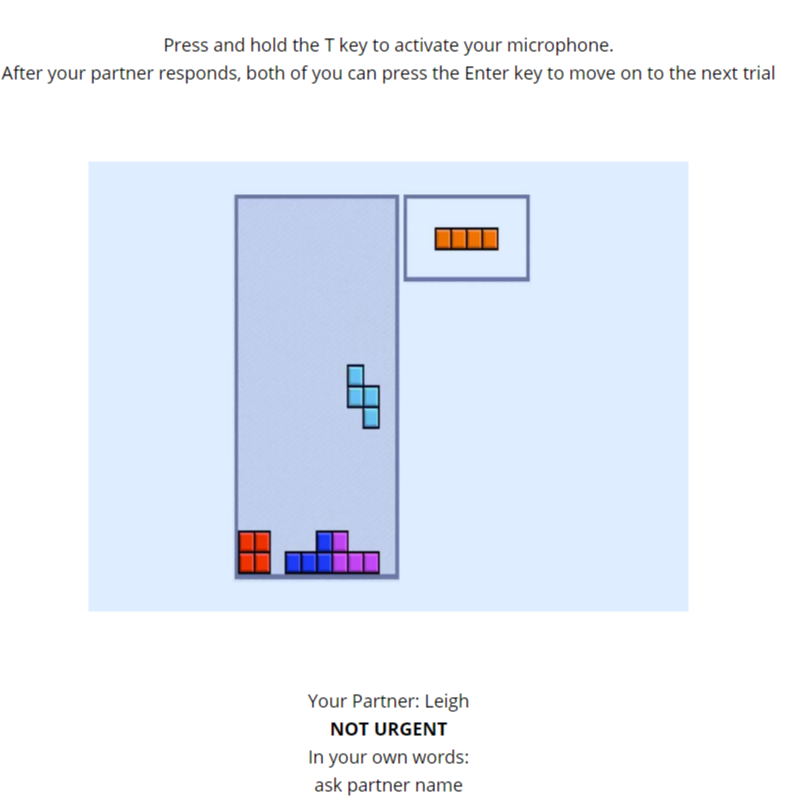}
  \caption{An example of a non-urgent trial that participants saw as a practice trial}
  \label{fig:nonurgent}
  \Description{A game of Tetris. Text beneath says "Non Urgent" in bold font on one line, "In your own words:" on the next line, and "ask partner name" on the next line.}
\end{figure*}

\begin{figure*}[h]
  \centering
  \includegraphics[width=0.5\linewidth]{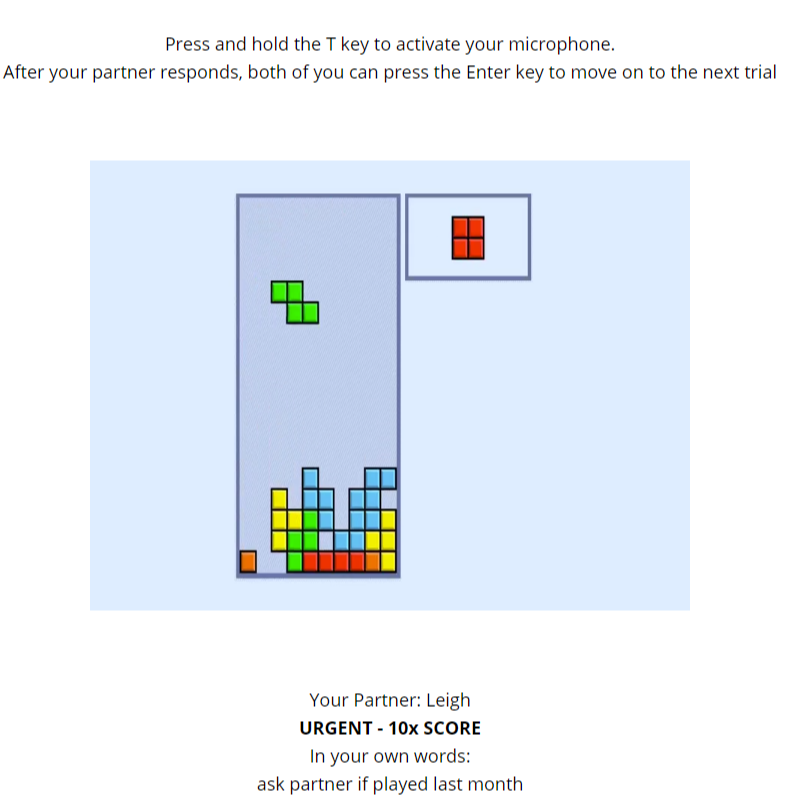}
  \caption{An example of an urgent trial that participants saw as a practice trial}
  \label{fig:urgent}
  \Description{A game of Tetris. Text beneath says "Urgent - 10x Score" in bold font on one line, "In your own words:" on the next line, and "ask partner if played last month" on the next line.}
\end{figure*}

The experiment followed a one-way within-subjects design. Interruption urgency was manipulated across two conditions: Urgent vs Non-Urgent. Following \cite{landesberger_what_2020}, urgency was manipulated by informing participants on urgent interrupting tasks (50\% of the trials) that their partner’s rating of their performance had a greater impact on their final score by a factor of 10 than the same ratings on non-urgent tasks. Interrupting tasks within the trials were either labelled preceding the interruption prompt as urgent- 10x score or not urgent (see Figures \ref{fig:nonurgent} and \ref{fig:urgent}). In this way, urgency was operationally defined as the interrupter's perceived cost of interrupting. This operationalisation ensured that urgency was defined explicitly to participants rather than being inferred by message content or confounded with interruption relevance.

\subsection{Measures}
\subsubsection{Interruption Onset}
The time it took for someone to commence an interruption (in milliseconds) was measured as the time from the interruption prompt being displayed to the moment the participant began their interrupting utterance. Distinct sounds were labeled automatically in all participant audio, with sound being defined as periods of noise louder than -40db (40db quieter than digital maximum for the recording) and sounds were separated when intervening silence lasted longer than 100ms. The lead experimenter then manually checked these sounds to ensure measurement accuracy and to identify the sounds that comprise the interruption utterance (i.e.  the interruption message and any preceding access rituals) in order to correctly identify the start of the interruption.

\subsubsection{Interruption Duration}
The lead experimenter also used these labeled sounds to identify the total length of time of the interruption (in milliseconds), measured from the interruption onset to the completion of the interrupting utterance. 

\subsubsection{Access Ritual Frequency}
Based on previous approaches \cite{krivonos_initiating_1975}, the lead experimenter categorized the types or access rituals used by participants to interrupt the Tetris player. Audio of participants’ verbal responses were used by the experimenter to determine whether each of the access ritual behaviors listed was present in the interruption. This included: Reference to other (i.e., Use of name or impersonal address); Apologizers (e.g., saying “sorry” or “excuse me”); Greeting (e.g., saying “hey”, “hi”); Filled openings (e.g., hesitations, disfluencies, “um”, “uh”, “hmm”, occurring at the beginning of an interruption) or Filled pauses (e.g., hesitations, disfluencies, “um”, “uh”, “hmm”, occurring elsewhere in an interruption). The presence of these were coded to produce a binary variable (1= access ritual present; 0= access ritual absent). 

\subsubsection{Open Ended Questions}
To gather further context and  gain an insight into the interruption strategies used, participants were asked four open-ended questions at the end of the experiment. Reflecting on the urgent and the non-urgent trials separately, participants were asked “how did you decide when to deliver messages to your partner?” and “how did you decide what to say to your partner?”

\subsubsection{Demographic Questionnaire}
Participants were asked a number of questions about themselves such as age, gender, and level of education, their level of experience with Tetris, and whether they believed their partner in the experiment to be another person playing live, a recording of a person, or a computer. 

\subsection{Procedure}
Participants were given information about the aims of the research, the data to be collected, and their data processing rights. Participants were then asked to give consent to take part in the study. Participants then were briefed on the procedure of the experimental task and told that they were being matched with a partner from an online Tetris website. They were also told that their performance would be rated by their partner and these ratings would determine which participant received a bonus prize. 

After an arbitrary delay, participants were told they had been connected to their partner and were shown generic partner information, including a unisex first name, a country of residence (e.g., “Leigh”, “Ireland” ) and some statistics indicating that their partner is a regular Tetris player (e.g. “11 hours played this month”). Next, the participants experienced two practice trial tasks, one non-urgent and one urgent. After completing each practice trial, the participant saw a screen for a random interval between 2500 and 3500ms informing them that their partner was rating their interruption. Next, participants were instructed that they would engage in 16 trials, after each of which their partner would rate their interruption. The experiment consisted of 16 Tetris trials and 16 interruption prompts. Each interruption prompt was presented only once to each participant. These were ordered randomly, with 8 prompts randomly assigned to each urgency condition across the 16 Tetris trials. The rating screen appeared for 2500 to 3500 ms after each trial. After all trials were completed, participants were asked to complete a brief questionnaire about their own background and their experience with the experiment, comprising the demographic questions and the open ended questions listed above. After completing the questionnaire, participants were fully debriefed explaining that their partner was actually a recorded member of the research team and that their performance was not being rated. They were informed that they were  eligible to receive a bonus prize, but this prize would be awarded randomly through selection of an anonymous Amazon Mechanical Turk ID. Participants were finally thanked for taking part and given instructions on receiving their payment.

\section{Results}
\subsection{Observed interruption behaviors}
\subsubsection{Quantitative Data Cleaning and Analysis Approach}
A total of 832 trials were recorded across the experiment. Trials in which technical issues rendered audio inaudible (N = 97 trials) or that were classed as extreme values  within the measures (+ or - 3 standard deviations from the mean;  N = 26 trials) were removed from the dataset. This resulted in a total of  709 trials by 46 participants being included in the final dataset for analysis. 

Linear mixed effects models were used to analyze the effect of urgency on interruption onset and interruption duration. Logit mixed effects models were used to analyze the effect of urgency on use of access rituals.  Mixed effects models are extensions of regression that allow data with hierarchical structures to be modeled in a way that accounts for both fixed effects of independent variables as well as participant-level and item-level effects through random intercepts and differences in magnitude of fixed effects through random slopes \cite{baayen_mixed-effects_2008, barr_random_2013}. Models were fit using the lme4 package version 1.1-26 \cite{bates_fitting_2015} in R version 4.0.3 \cite{r_core_team_r_2020}. Following best practices, we started with the maximal random effect structure for the experiment (e.g. random slopes and intercepts at the subject- and item-level) and incrementally reduced complexity for a given model until models could converge \cite{barr_random_2013}. To improve reproducibility, full model syntax and random effect outputs are included in supplementary materials for each model \cite{meteyard_best_2020}. 

\subsubsection{Interruption Onset (H1)}
\begin{figure*}[h]
  \centering
  \includegraphics[width=0.75\linewidth]{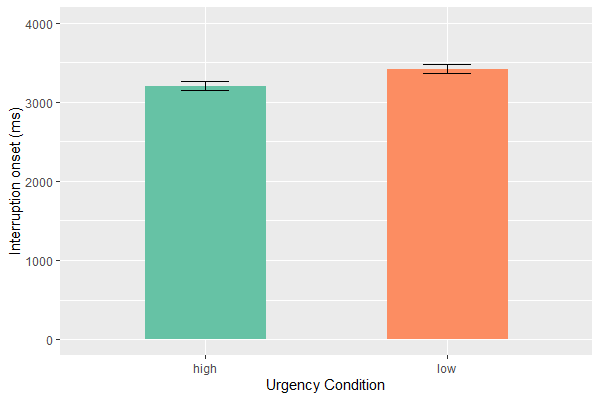}
  \caption{Means and and standard errors for interruption onset times by urgency condition}
  \label{fig:graph}
  \Description{A bargraph with two bars including error bars. The bar labeled low for the x-axis of urgency condition is taller than the bar labeled high on that axis. The error bars on the two bars do not overlap.}
\end{figure*}

We found a statistically significant effect of urgency [Unstandardized $\beta$ =23.83, SE $\beta$ =112.58, 95\% CI [7.45, 458.30], t=-2.07, p=.04] with participants delaying significantly longer before non-urgent  interruptions (M = 3419ms; SD = 1312ms) as compared to urgent interruptions (M = 3200ms; SD = 1276ms). This supports H1 and is visualized in Figure \ref{fig:graph}. Descriptive statistics for interruption onsets overall and by condition are reported in Table \ref{tab:descr}.

\subsubsection{Interruption Duration (H2)} 
We found no statistically significant effect of urgency  [Unstandardized $\beta$=32.25, SE $\beta$=37.10, 95\% CI [-40.57, 105.07], t=-0.87, p=.39] on the duration of interruption. This means that H2 was not supported. Descriptive statistics for interruption durations overall and by condition are reported in Table \ref{tab:descr}.

\begin{table}
\centering
\caption{Table of means and standard deviations for interruption onset and interruption delay by urgency condition.}
\label{tab:descr}
\begin{tabular}{|c|c|c|c|} 
\hline
Measure & Urgency condition & Mean (ms) & SD (ms) \\ 
\hline
\multirow{3}{*}{\begin{tabular}[c]{@{}c@{}}Interruption \\Onset\end{tabular}} & High & 3200 & 1227 \\ 
\cline{2-4}
 & Low & 3419 & 1311 \\ 
\cline{2-4}
 & Overall & 3293 & 1699 \\ 
\hline
\multirow{3}{*}{\begin{tabular}[c]{@{}c@{}}Interruption \\Duration\end{tabular}} & High & 1400 & 288 \\ 
\cline{2-4}
 & Low & 1431 & 299 \\ 
\cline{2-4}
 & Overall & 1419 & 550 \\
\hline
\end{tabular}
\end{table}

\subsubsection{Access Rituals (H3)}
We found  no statistically significant effect of urgency  [Unstandardized $\beta$=-0.20, SE $\beta$=0.29, 95\% CI [-0.77,0.37], z=-0.69, p=.49] on the likelihood of using access rituals in interrupting utterances. This means that H3 was not supported. Across the data, 23 out of 46 participants used no access rituals at all, with four participants using access rituals on more than half of their trials. Descriptive statistics for counts of access ritual behaviors overall and by condition are reported in Table \ref{tab:rituals}.

\begin{table}[hbt!]
\caption{Table of counts of trials containing access rituals by urgency condition.}
\label{tab:rituals}
\begin{tabular}{|c|c|c|}
\hline
 & \textbf{\begin{tabular}[c]{@{}c@{}}Trials containing \\ an access ritual\end{tabular}} & \textbf{\begin{tabular}[c]{@{}c@{}}Trials without \\ an access ritual\end{tabular}} \\ \hline
\textbf{High} & 57 &  295 \\ \hline
\textbf{Low} & 51 &  306 \\ \hline
\textbf{Overall} & 108 &  601 \\ \hline
\end{tabular}
\end{table}

\subsection{Self-identified Interruption Strategies}
\subsubsection{Data Analysis Approach}
Answers to open-ended questions were analyzed through thematic analysis by the lead author (who has experience conducting qualitative analysis and has a background in interruptions and speech interface research), using a hybrid approach \cite{fereday_demonstrating_2006}. Initial codes were generated inductively, guided by prior work on interruptions and speech, with themes also developed deductively through a staged review of the data and initial codes, consistent with a reflexive approach to thematic analysis \cite{braun_using_2006}. For the questions regarding timing, initial codes were generated to reflect literature on speed-accuracy tradeoffs for interruptions \cite{brumby_fast_2011}, with timing strategies coded as focusing on either the speed of the interruption, accuracy in the interrupting task (i.e. avoidance of error in talking to one’s partner), or the accuracy of the primary (Tetris) task.  A third code represented responses that gave no indication of a conscious strategy. Note that time spent on the primary task is a direct function of the speed of the interrupting task, in that both tasks end when the interrupting task is completed, so speed of the primary task was not an initial code. For questions regarding what participants said to their partner, initial codes were generated to reflect literature on urgent speech \cite{landesberger_urgent_2020, hellier_perceived_2002}, with speaking strategies coded as phrasing (semantic characteristics) or delivery style (prosodic characteristics). A third code represented responses that gave no indication of a strategy. Because of the hybrid approach used in our thematic analysis \cite{fereday_demonstrating_2006}, these inductive codes served as a starting point and do not encompass all of the final themes which we generated deductively through staged review. 

\subsubsection{Interruption Timing Strategies}
Four themes for interruption timing strategies were generated inductively. Participants felt they either timed their interruption in a way that always prioritized accuracy, in a way that always prioritized speed, mixed strategies according to characteristics of the interrupting task (i.e. interrupting message content), or mixed strategies according to characteristics of the Tetris task. Themes are presented below along with counts of how many participants in each condition mentioned a given strategy (out of a total of 52 participants).
\\ \\
\textit{Prioritizing Speed (Non-urgent: 9 participants, Urgent: 30 participants)} \\
Many participants stated that, when completing the trials, they interrupted as soon as they could.  This strategy was mentioned more frequently when discussing strategies in the urgent trials, although it was mentioned when discussing non-urgent trials too. Some participants did not consider the state of the Tetris task when planning their interruption  stating that \textit{“[I interrupted] as soon as possible, the timing of Tetris didn't occur to me”} (P09) while other explanations were more brief, stating they interrupted \textit{“as soon as I could”}, \textit{“as soon as possible”}, or \textit{“as soon as they appeared”} (Ps 02, 09, 41). The difference in prevalence of the speed strategy between conditions supports the quantitative results highlighting faster interruption onset in the urgent trials.  \\ \\
\textit{Prioritizing Accuracy (Non-urgent: 6, Urgent: 0)} \\
Especially when discussing the non-urgent condition, participants mentioned the importance of accuracy, trying to prevent errors in interruption delivery, sacrificing speed. Some participants specifically mentioned sacrificing speed across the entirety of a condition, as opposed to timing interruptions based on features of the Tetris task or of the interrupting task. 
\begin{center}
\textit{“[I] Took my time deciding on how to word and when to deliver the question”} (P28) \\ 
\textit{“[I] just decided to say it casually. not make him feel like he needs to answer too quickly for the low urgency trials.” (P44)} \\
\end{center}
The mention of taking one’s time in non-urgent trials but not in urgent trials is somewhat surprising, as past research has indicated that people generally prefer to interrupt as quickly as possible when not specifically instructed otherwise \cite{brumby_recovering_2013, horrey_driver-initiated_2009}. It may be that participants saw this strategy as more appropriate, but not well-suited to urgent interruptions, and thus were more likely to use this strategy in non-urgent trials. Again this supports our quantitative findings of taking longer to start an interruption in non-urgent trials than urgent trials. \\ \\
\textit{Tetris Task Characteristics (Non-urgent: 33, Urgent: 18)} \\
Fifty-one responses mentioned the importance of using characteristics of the Tetris task to decide when to interrupt. From the comments some participants describe themselves as being sensitive to subtask boundaries (Non-urgent: 6, Urgent: 3), to the player’s cognitive load (Non-urgent: 25, Urgent: 14), or mention the Tetris task without specifying the characteristics of the task they were sensitive to (Non-urgent: 18, Urgent: 1). 

\begin{figure}[h]
  \centering
  \includegraphics[width=0.75\linewidth]{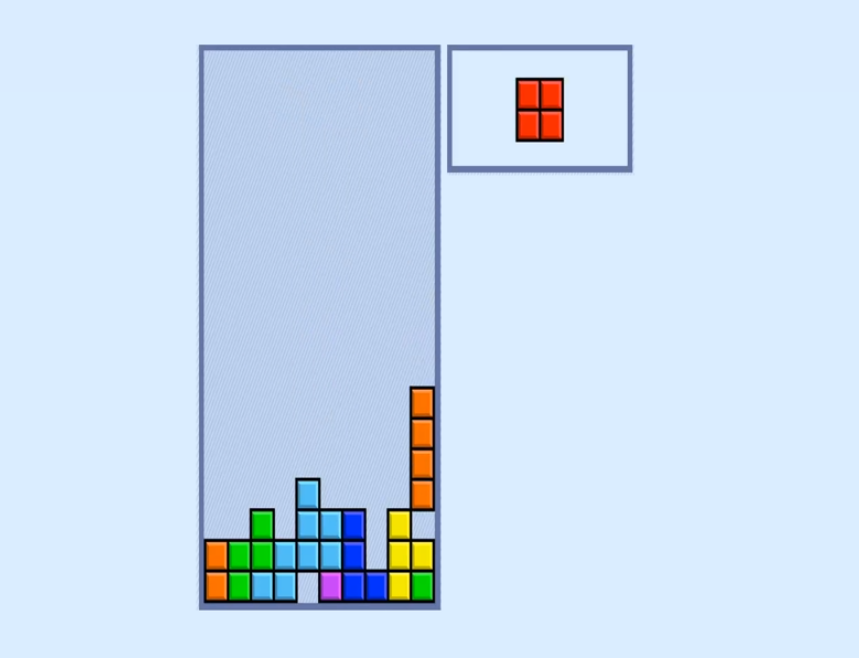}
  \caption{An example of a possible subtask boundary which some participants identified as a good moment to interrupt. Note that the orange Tetris block on the far right is the currently falling block.}
  \label{fig:breakpoint}
  \Description{A game of Tetris. An orange block is one space above a gap where it will presumably land.}
\end{figure}

Those who mentioned subtask boundaries as a cue for timing their interruptions seemed to plan interruptions for when a Tetris piece was in its final destination or at the top of the screen - when the subtask of placing a piece had just finished and the next subtask was just beginning (see Figure \ref{fig:breakpoint}). They tend to emphasize that they would interrupt \textit{“When there was a new block so that it was at the top of the screen”} (P10) or \textit{“As soon as a block was placed and a new one was at the top of the screen”} (P12). 

There were also those that attempted to identify moments in which their partner was  under less cognitive load, unburdened by making a decision for the Tetris task. They focused on moments when \textit{“placing a block was not too difficult”} (P12) or when \textit{“the game was not intense.”} (P25) as well as opportune moments when the participants perceived that the player had clearly finished making a decision \textit{“I delivered when I felt she had selected a spot for the falling piece.”} (P29)

Others were less specific about the characteristics of the game they prioritized but still indicated that they used the Tetris task state to assess when was the right time to ask a question: \textit{“I watched the play and then asked the question”} (P01).

There is likely considerable overlap in Tetris task-dependent reasons that these participants picked their moments to interrupt. Natural breakpoints such as subtask boundaries are frequently the lowest cognitive load moment within a task and are thus ideal for interruptions \cite{brumby_recovering_2013, bailey_understanding_2008}. Choosing subtask boundaries as moments of interruption may well be seen as selecting the moments they find to be the least intense or the most convenient. Likewise, selecting moments between decisions construes the game of Tetris as made up of a series of decisions at subtasks. We therefore propose these descriptions of Tetris-task dependent strategies fit together in the same theme. \\ \\
\textit{Message Content  (Non-urgent: 2, Urgent: 0)}\\
One relatively rare strategy was to time interruptions depending on the content of that interruption.  Two participants mentioned that the timing of their utterances depended on what question they were asking their partner. One of these participants explained their exact rationale, saying \textit{“I tried to wait until a piece had been played if it was a longer question, if it was a simple and short question I asked it straight away”} (P51) indicating that the message content was a primary strategy selection criteria, selecting the Tetris task strategy for long questions and the speed strategy for short questions.\\ \\
\textit{No Strategy (Non-urgent: 2, Urgent: 4)}\\
Some participants either explicitly noted that they did not think about how to time their interruptions and as such identified no strategy, suggesting that they \textit{“didn't really change [their] communication one way or the other.”} (P21).

\subsubsection{Interruption Structure}
For the questions regarding what participants said to their partner, three clear themes were generated inductively. Participants either focused primarily on the way they phrased their message (i.e. word choice), they focused on how delivered their message (i.e. prosodic features), or they mixed strategies according to the characteristics of the interrupting task (i.e. interrupting message content). These themes are explored below with comparisons of frequency in the non-urgent and urgent conditions.\\ \\ 
\textit{Phrasing (Non-urgent: 36, Urgent: 33)} \\
A major theme in how participants structured their interruptions was phrasing. Within this theme, three strategies were identified, delineating what characteristic of their phrasing participants prioritized: word length (Non-urgent: 18, Urgent: 21), naturalness (Non-urgent: 16  Urgent: 9), or other (Non-urgent: 2, Urgent: 3). 

Many participants who focused on the phrasing of their interruptions did so by trying to interrupt with as few words as possible, sometimes explicitly acknowledging that this was to reduce cognitive load on their partner:  \textit{“I used as few words as possible, so she didn't have to think about it”} (P15). Others who focused on word length took the opposite approach, seeking to avoid error by \textit{“ask[ing] questions elaborately”} (P01), specifying that they \textit{“Said it in detail so he would give me the correct answer.”} (P44). This phrasing strategy was less prevalent than the former, but both were distributed similarly across urgency conditions.

For some, phrasing was not primarily about length, but about asking questions \textit{“that made sense”} (P42), that were phrased as \textit{“the questions I would normally ask an acquaintance.”} (P23), and questions that \textit{“reflect what needs to be asked.”} (P47). It isn’t clear whether participants perceived natural phrasing as consistent with shorter phrases, longer phrases, or neither, so these strategies were grouped together under the theme of phrasing. There were also participants who prioritized other ways of phrasing such as using \textit{“the most informative way to ask the question.”} (P40). These diverse strategies around phrasing were classified as part of the same broader phrasing theme. \\ \\
\textit{Delivery (Non-urgent: 5, Urgent: 11)} \\
Another major theme in how participants structured their interruptions was delivery, focusing in particular on prosody - the way their speech sounded. This theme includes three strategies concerning delivery, each delineated by which characteristic of their their delivery participants mentioned: tone (Non-urgent: 1, Urgent: 1), clarity (Non-urgent: 4, Urgent: 4), or speed (Non-urgent: 0, Urgent: 6).

One participant focused on their tone of voice, seeking to deliver interruptions in \textit{“a calm voice to not startle my partner”} (P24), using this strategy in both urgency conditions: \textit{“Again, I said it calmly”} (P24).

Others who focused on delivery instead prioritized clarity, seeking to deliver interruptions \textit{“clearly so she can understand.”} (P47). These participants mention focusing less on choosing their words, instead ensuring that they \textit{“spoke it clearly.”} (P45).

A focus on clarity did not always pay off however, as one participant using this strategy expressed regret for not instead focusing on phrasing.

\begin{center}
\textit{“I tried to make my questions as clear as possible, but in hindsight I think I probably should've made an effort to make my questions shorter as though I started when I thought it was a good time to talk, actually by the time I'd finished asking and it was time for her response it was in the middle of what I'd consider a high risk moment in the game!”} (P16)    
\end{center}

This expression of regret gives insight into the extent to which themes overlapped and the dynamic nature of strategy selection. Finally, some participants mentioned that they  \textit{“tried to speak quickly”} (P29). It should be noted that speaking quickly was considered a delivery strategy in this analysis, but it may be highly correlated or conflated with the strategy of minimising phrase length for individual participants, as mentions of speaking speed were typically short vague expressions like \textit{“I spoke quicker”} (P30). \\ \\
\textit{Message content (Non-urgent: 5, Urgent: 4)}\\
Some participants mentioned varying their strategies for structuring interruption \textit{“based on the type of question.”} (P13). Participants who varied strategies did not give much indication of which features of the content of the message were relevant to them nor how they varied their strategy, vaguely alluding to how they \textit{“relied more on the text that was at the bottom of the screen”} (P03) in one urgency condition or the other. This theme may not lend much insight to how message content impacts strategy selection, but it nonetheless provides some evidence that message content may impact strategy selection for some people, and that strategies are not rigid functions of urgency or individual preferences.\\ \\
\textit{No strategy (Non-urgent: 6, Urgent: 4)}\\
Just as was the case with timing strategies, some participants either explicitly noted that they did not think about how to structure interruptions or gave short or vague responses like \textit{“[I] read the description and made a decision”} (P08) that did not fit into any of the above themes, or explicitly stated \textit{“I didn't really change my communication one way or the other.”} (P21).

As was the case with timing strategies, a lack of stated strategy is not necessarily an indication of no strategy. The above quote from P21 indicates that some participants may have thought about this question comparatively, noting whether their interruption differed between conditions but not explaining their strategy if it was consistent. Again, no participant in this theme indicated that they randomly altered their interruption structure or that they avoided using a consistent strategy, so this theme is best viewed as an absence of an explicit acknowledgement of a strategy rather than an absence of strategy per se.

\section{Discussion}
Building on recent work on the design of proactive speech agents \cite{cha_hello_2019}, our study aims to give insight into how interruptions should be designed, especially in contexts where interruptions may be urgent or time-sensitive. Our research, built around a new paradigm for eliciting speech interruptions in a dual-task context, illuminates the variety of strategies that people employ when interrupting people who are engaged in another task. These strategies could be adopted by speech agents. Through our mixed-methods study we find that people tend to interrupt people significantly sooner when delivering an urgent interruption than when the interruption is non-urgent. That said, there are many different types of perceived strategies taken by people who are looking to interrupt, highlighting the critical  contribution of individual differences to interactions. We found that some participants identify their strategies for timing interruptions as being based on characteristics of either the interruption itself or of the task they are interrupting, while others apply consistent strategies irrespective of the nature of a task. We also found that participants identify their strategies for structuring interruptions as particularly focused on word length, utterance naturalness, clarity, and tone. Below, we discuss these findings in the context of the interruptions literature and the design of proactive speech agents. 

\subsection{Interruption Strategies are Highly Diverse}
Through thematic analysis of participants’ descriptions of their strategies, we have gained some key insights into how spoken interruptions are timed and structured. While some people use characteristics of their partner’s primary task (Tetris) to determine when to interrupt, others use characteristics of the interrupting message or interrupt according to fixed strategies irrespective of the tasks. This is consistent with other work on multitasking that found a similar complex mix of strategies for self-interruptions \cite{cha_hello_2019, dabbish_why_2011}. As modeling complex situations like driving or daily life is still an ongoing challenge \cite{semmens_is_2019, cha_hello_2019}, the insight we provide about the diversity of strategies people use to time interruptions should help to guide speech agent design as task modelling capabilities improve. Future work should investigate whether moments that interrupters identify as natural breakpoints (e.g., when a Tetris piece is at the bottom of the screen) correspond with when they interrupt people. This work would help unite existing understandings of natural breakpoints \cite{borst_problem_2010, janssen_natural_2012} with the ongoing work on communication during multitasking. Furthermore, future work may consider whether an interrupter’s expertise in a primary task influences perception of breakpoints and thus impact interruption strategies. This may be particularly important for increasingly complex tasks like driving or workplace environments in which task understanding requires greater expertise than does Tetris.

Themes regarding the structure of interruptions unite present knowledge of urgent speech \cite{landesberger_urgent_2020, hellier_perceived_2002} with our understanding of explicit goals in multitasking \cite{brumby_recovering_2013}, indicating that people alter both their word choice and their prosody depending on the urgency of an interruption. Speech agent designers could implement this feature of human speech production into synthesized speech, allowing users to hear particular notifications in an urgent voice while using a non-urgent voice for other notifications. Recent work has begun to explore this approach, finding that the use of more assertive voices significantly impact the speed of task switching from a complex primary task \cite{wong_voices_2019}. From our findings, it is important to consider that the speech properties people used to communicate urgency varied. Future work should investigate if preferences of expressions of urgency used by speech agents likewise vary between individuals.

\subsection{Few People Use Access Rituals}
This work sought to investigate the use of access rituals - short verbal behaviors that signal a request for a listener’s attention -  in spoken interruptions. Not much is known about how people initiate spoken interruptions, so it was unknown whether people used access rituals at all when interrupting. We found that urgency did not influence access ritual use. Most participants did not use them across the trials, yet some frequently did. The reason for this is unclear.  People may have felt they already had social access to their partner due to both taking part in an experiment, and thus did not need to request it. It may also be that the relative importance of interrupting was so high as to diminish the social need for access rituals, or that there is a natural variability in the use of access rituals across the population observed here compared to that in the original research (i.e. American college students who were previously acquainted and interacting face to face) \cite{krivonos_initiating_1975}. Nonetheless, that some participants did use access rituals frequently may be of interest to speech agent designers. Future work should investigate whether the use of access rituals by nonhuman agents is preferred by some users or if, like other humanlike personalizations to agents, this is seen as unnatural, fake or unpleasant \cite{clark_what_2019, doyle_mapping_2019}.

\subsection{Urgent Interruptions Are Delivered Sooner Than Non-urgent Interruptions}
Quantitative findings regarding people’s interruptions indicate that urgent interruptions are initiated more quickly  than non-urgent interruptions, but they are not different in duration. Urgent interruptions having shorter delays is in line with previous findings in which people prioritize an interrupting task over a primary task when told to do so \cite{brumby_recovering_2013}. While the size of the effect of urgency on interruption onset was small, seminal work on interactive behaviour highlights the importance of small differences in time measurements \cite{gray_milliseconds_2000}. These can reveal user microstrategies that can inform interactive system design \cite{gray_milliseconds_2000}. Our qualitative findings support the notion that users prioritized speed in urgent trials, indicating real strategy differences in interruptions according to urgency. Indeed, in contexts where stakes are higher (e.g., driving) or where task states are more difficult to assess, quantitative differences of the size found in our study may in fact be critical, and effects in such contexts may become even larger. Interruptions were quantitatively and qualitatively different depending on the level of urgency, indicating that the paradigm successfully elicited utterances that differed in urgency. That interruptions did not differ significantly in duration, contradicting the theme of speaking faster and using shorter utterances for urgent interruptions, may reflect the relatively minor impact of both prosodic and semantic adaptations to urgency. While work has begun on identifying the prosodic features of urgent speech \cite{landesberger_urgent_2020}, more work is needed to further investigate the magnitude of the effect of urgency.

\subsection{A New Paradigm for Dialogue Interruptions Research}
While interruption properties are well-studied, communication in multitasking environments like this is not. The proposed experimental paradigm represents a first step in better understanding this communication. This work further sought to explore the importance of characteristics of the interrupting message, in this case urgency, and characteristics of a partner’s primary task in shaping communication strategies. The paradigm proposed here uses a gamified approach like other recent work in eliciting human-speech in the design of agent speech \cite{landesberger_urgent_2020, landesberger_what_2020}, but it is flexible to different primary tasks and different independent variables. Furthermore, the elicitation paradigm was useful in generating speech that was meaningfully impacted by the independent variable of interest (urgency) with crowdworkers as participants. This feature should help researchers in this area obtain larger and more diverse samples in order to inform speech agent design. 

\subsection{Limitations}
While this work focuses on initiating conversation with people actively engaged in another task, not all agent-initiated interruptions will need a response. Indeed, many interruptions that occur during complex, continuous tasks include information delivery rather than requests of information from the user (e.g. navigation information while driving).  Insights from this work may improve the design of interruptions that require a spoken response from users, but they may not be applicable to other interrupting contexts. Likewise, this work looks at the interruption of a low-risk task, and interruption strategies may be more divergent or entirely different for contexts in which errors are more costly. While our results illustrate a complex assortment of interrupting strategies, these emerged from a constrained continuous task and simple interrupting utterances. This work serves as an early step in understanding how agents might coordinate interruptions that vary across dimensions beyond just urgency and in contexts more difficult to model than Tetris. Designing for real world interactions of this sort will require much further work. Urgency in this study was operationally defined as a reflection of how harshly disruptiveness to the partner’s primary task (Tetris) would be judged by their perceived partner. Participants may instead have interpreted urgency as indicative that interruptions are time sensitive or that errors during interruption were more costly. In this way, the subtle ambiguity about the meaning of urgency may limit generalizability across other contexts of urgency. Finally, participants in this study were crowdworkers interacting with recordings of people rather than dyads of people interacting online or while physically copresent. More work is needed to investigate how social dynamics such as personal relationships between people or physical copresence affect the ways people interrupt others who are engaged in another task. 

\section{Conclusion}
This work aims to serve as a first step toward greater understanding of spoken interruptions of complex, continuous tasks for the purpose of engaging in conversation. As speech agents are embedded into more of the technology around us, the design of spoken interruptions grows increasingly important. The gamified paradigm demonstrated here allows designers to understand spoken interruptions in general and to tailor those interruptions to a variety of primary tasks, interruption content, and variables of interest. We hope to empower speech agent designers to quickly and easily gather data about how people interrupt those engaged in another task, as we see this as a critical question for the future of proactive speech agent development.

\begin{acks}
This research was conducted with the financial support of the ADAPT SFI Research Centre at University College Dublin. The ADAPT SFI Centre for Digital Content Technology is funded by Science Foundation Ireland through the SFI Research Centres Programme and is co-funded under the European Regional Development Fund (ERDF) through Grant \# 13/RC/2106\textunderscore{}P2.
\end{acks}

\bibliographystyle{ACM-Reference-Format}
\bibliography{Tetris}


\end{document}